\documentclass[12pt,preprint]{aastex}
\usepackage{times}
\usepackage{amsmath}
\usepackage[dvipsnames]{xcolor}

\shorttitle{A possible selection rule for flares causing sunquakes}
\shortauthors{Chen \& Zhao}

\begin{document}

\title{A Possible Selection Rule for Flares Causing Sunquakes}

\author{Ruizhu Chen\altaffilmark{1} and Junwei Zhao\altaffilmark{1}}
\altaffiltext{1}{W.~W.~Hansen Experimental Physics Laboratory, Stanford
University, Stanford, CA 94305-4085}

\begin{abstract}
Sunquakes are helioseismic power enhancements initiated by solar flares, but not all flares generate sunquakes. It is curious why some flares cause sunquakes while others do not. Here we propose a hypothesis to explain the disproportionate occurrence of sunquakes: during a flare's impulsive phase when the flare's impulse acts upon the photosphere, delivered by shock waves, energetic particles from higher atmosphere, or by downward Lorentz Force, a sunquake tends to occur if the background oscillation at the flare footpoint happens to oscillate downward in the same direction with the  impulse from above. To verify this hypothesis, we select 60 strong flares in Solar Cycle 24, and examine the background oscillatory velocity at the sunquake sources during the flares' impulsive phases. Since the Doppler velocity observations at sunquake sources are usually corrupted during the flares, we reconstruct the oscillatory velocity in the flare sites using helioseismic holography method with an observation-based Green's function. A total of 24 flares are found to be sunquake active, giving a total of 41 sunquakes. It is also found that in $3-5$ mHz frequency band, 25 out of 31 sunquakes show net downward oscillatory velocities during the flares' impulsive phases, and in $5-7$ mHz frequency band, 33 out of 38 sunquakes show net downward velocities. These results support the hypothesis that a sunquake more likely occurs when a flare impacts a photospheric area with a downward background oscillation.

\end{abstract}



\section{Introduction}
\label{sec1}
Sunquakes are photospheric oscillatory power enhancements initiated by  
{solar flares}, exhibiting as photospheric helioseismic waves  
expanding from the flaring sites visible about {10-30} min after the flare onset. 
The flare-excited helioseismic waves  penetrate into the Sun, travel in the solar 
interior, and get reflected back to the photosphere where they are observed as ``ripples". 
Sunquakes were first predicted by \citet{Wol72} as a 
consequence of energy release in major solar flares, while the first observation of such events was 
made decades later by \citet{Kos98} using data from the {\it Solar and Heliospheric 
Observatory} / Michelson Doppler Imager \citep[{\it SOHO}/MDI;][]{Scher95}. 
Following this discovery, more sunquake events were reported  
\citep[e.g.,][]{Don05, Kos06, Kos07, Zkv11}, 
and more recently, {responses} in the magnetic field and in the chromosphere {to sunquakes}
were also discovered \citep{Zha18,Qui19}. But the
occurrence of sunquakes is still relatively rare {---} only a
fraction of strong solar flares are accompanied by detectable
sunquakes, {and weak flares cause sunquakes with an even lower rate} \citep{Bui15, Sha20}. 
{The low occurrence rate makes} the understanding of the sunquake-triggering mechanism difficult. 

Mechanisms that are proposed to trigger sunquakes can be 
divided into two categories. In the first category, sunquakes are excited by sudden 
photospheric pressure perturbations caused by shock waves, photospheric 
heating, or Alfv\'en-wave heating. Downward-propagating hydrodynamic shock 
waves can be generated through thick-target heating of the 
chromosphere by nonthermal electron \citep{Kos98, Kos07}, or driven by energetic proton 
beams \citep{Zha07}. These mechanisms are supported by hard 
X-ray emissions \citep{Don05, Kos07} and gamma-ray emission \citep{Zha07} 
that are cospatial with sunquakes.  
Photospheric heating can be caused by backwarming \citep{Don06, Lin08}, 
where particle beams heat the chromosphere and chromospheric radiation heats the 
photosphere. This mechanism is supported by white light emissions \citep{Don06,Bui15} cospatial with some 
 sunquakes. Downward-propagating Alfv\'en waves can heat the chromosphere similarly 
 to electron beams and produce explosive evaporation \citep{Rus13, Ree16}, 
 possibly contributing to sunquake excitation, as indicated by the observation by \citet{Mat15}.
In the other category, Lorentz force is proposed to act as a sunquake driver 
\citep{Hud08,Fis12}. Permanent changes in the photospheric magnetic field that often 
occur during flares \citep{Wang94, Kos01, Sud05,Pet10,Wang10,Sun17} cause Lorentz-force 
changes in the photosphere, hence a pressure perturbation that can excite sunquakes.  
In particular, since the magnetic field in the photosphere is expected to be more 
horizontal after magnetic eruption, the change of Lorentz force is expected to be 
predominantly downward acting on the photosphere and the solar interior \citep{Fis12}. 
This scenario is also supported by observations \citep {Wang10, Pet10, Pet13}. 
 
However, none of these proposed mechanisms satisfactorily explain the disproportion 
between the occurrence rates of strong solar flares and sunquakes --- although 
sunquakes tend to occur during strong flares, only a fraction of strong flares cause sunquakes 
and {an even smaller fraction of weak flares cause sunquakes}..
Trying to explain this disproportion, in this paper we explore  
certain conditions on the background oscillatory velocity during the impulsive phases of the sunquake-generating 
flares, and propose a selection rule for the sunquake occurrences. 
As is known, background oscillations are stochastically excited by solar convection and 
are replete in both quiet-Sun regions and active regions. Is it possible that 
sunquakes are actually background helioseismic oscillations that happen 
to be enhanced by solar flares? A possible scenario is that, during a flare's impulsive 
phase, when the pressure perturbation due to shock waves or energetic particles reach 
the photosphere from the higher atmosphere, or when the downward Lorentz impulse acts on the photosphere, 
the background oscillation at the flare footpoints happens to oscillate downward 
in the same direction with the perturbations 
and thus gets enhanced to a level that is detectable from the ambient background oscillations and waves. 
Oppositely, an upward-oscillating background at the flare footpoints may fully or partially offset  
the momentum from the flare, decreasing the flare's probability of causing a sunquake. 
As a result, even strong flares may not generate a sunquake. 
{Meanwhile, it is worth pointing out that the sunquake-like events, caused by solar convection 
and whose ripple-like features are mostly lost in the ubiquitous oscillations, are also abundant 
in the quiet-Sun regions.}

To examine this hypothesis, we use high-resolution continuous observations from {\it Solar 
Dynamics Observatory} / Helioseismic and Magnetic Imager \citep[{\it SDO}/HMI;][]{Scher12, 
Schou12}, to perform a systematic survey on { flares stronger than M6.3 for a sample 
of $\sim30$ sunquake events} during Solar Cycle 24, 
searching for sunquakes and examining whether the oscillatory velocities at the sunquake sites 
are downward during the flares' impulsive phases. However, the direct Doppler-velocity 
observations in the flaring regions are often corrupted during the flares' impulsive phases, 
resulting from irregular spectrum lines 
due to the high temperature and violent dynamics, therefore the raw data are not immediately suitable 
for this analysis (see Section~\ref{sec_Data} for details). 
As an alternative, we reconstruct the sunquake oscillatory velocities at the flaring sites during the flare 
by employing the helioseismic holography method \citep{Lin97, Lin00}, which, in principle, reverses the 
expanding sunquake ripples backward to the sunquake initiation time and location 
by use of a Green's function. While previous holographic analyses usually employ 
a theoretical Green's function, we adopt an observation-based Green's function for a more precise 
temporal determination necessary for this study.  
In this paper, we introduce the analysis method in Section~\ref{sec_Method} and data in Section~\ref{sec_Data}. 
We show our analysis procedure and results on one sunquake example in Section 4 , and the 
overall statistics of the whole survey are presented in Section 5. We then discuss our results in Section 6.


\section{Method}
\label{sec_Method}

\subsection{Observation-Based Green's Function}
\label{sec_Green}

Green's function describes how waves propagate in space and time from an impulse at the source, and is used in the holography technique to reverse the waves backward in time and reconstruct the oscillatory signals at the source. Theoretical Green's functions, noise free and calculated from a quiet-Sun model, were used in previous studies \citep{Lin00}. However, theoretical Green's functions are biased in determining the sunquake initiation time due to the fact that the sunspot differs from the quiet-Sun model, and that waves starting from sunspot regions are known to have a shorter travel time than those starting from quiet-Sun regions \citep{Duv96}. Besides, the dispersion relation of acoustic waves, which influences travel times as well as amplitudes of waves, is different in active regions from that in quiet-Sun regions. Therefore, an observation-based Green's function for waves originating from active regions is more realistic for our analysis purpose, and is expected to give a better determination of starting time for sunquakes. 

We calculate such Green's functions using the Doppler-velocity data by use of the time--distance helioseismic method, and a conceptually similar but  technically different approach was developed before by \citet{zha13}.  Ideally, Green's functions should be calculated and employed case by case, i.e., for each sunspot and for various origin locations. However, it is computationally expensive to do so, and besides, observations of one sunspot are not large and long enough for averaging to obtain a less noisy Green's function.  Meanwhile, although sunspots vary in sizes and shapes, it is found that the most significant factor affecting the measured travel times is which part the waves originate from, i.e., umbra, penumbra, or quiet region around sunspot \citep{Duv96}. So, we build three sets of Green's functions for waves propagating outward from umbra, penumbra, and near-sunspot quiet regions, respectively, by averaging over all sunspots in this survey and over all pixels in each location category, neglecting variations of different active regions and variations of relative locations inside umbra, penumbra, and quiet regions. 

\begin{figure}[!t]
\epsscale{0.50}
\plotone{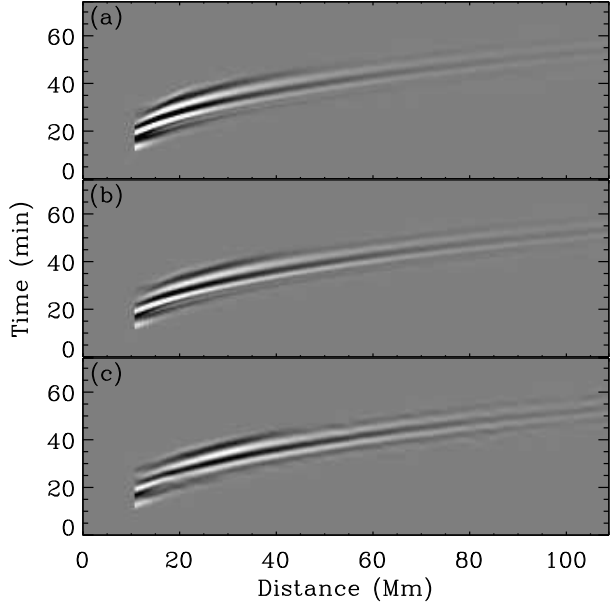}
\caption{Observation-based Green's functions (with a normalization factor $\alpha_{\bf r\rightarrow r'}$) for waves traveling out from (a) quiet regions, (b) penumbra, and (c) umbra and are received in quiet regions in the $2-7$ mHz frequency band. }
\label{GF_fig}
\end{figure}

The calculation of an observation-based Green's function uses the average cross-covariance function, $\Phi(|{\bf r}-{\bf r'}|,\tau)$, computed from time sequences $\Psi(t)$ at locations ${\bf r}$ and ${\bf r'}$,
\begin{equation}
\Phi(|{\bf r}-{\bf r'}|,\tau)=\langle\Psi_{{\bf r}}(t)\star\Psi_{{\bf r'}}(t)\rangle_{|{\bf r}-{\bf r'}|}=\langle\int\Psi_{{\bf r}}(t)\Psi_{{\bf r'}}(t+\tau){\mathrm d}t\ \rangle_{|{\bf r}-{\bf r'}|},
\label{cov}
\end{equation}
where $\star$ denotes cross-covariance calculation, and $\langle \  \rangle_{|{\bf r}-{\bf r'}|}$ denotes averaging over a large field for all ${\bf r}$ and ${\bf r'}$ with ${|{\bf r}-{\bf r'}|}$ equal to a certain value. In calculating $\Phi$, any pairs of ${\bf r}$ and ${\bf r'}$ can be treated as a source and a receiver, and vice versa. In observations, $\Psi_{\bf r}(t)$ and $\Psi_{\bf r'}(t)$ both contain plenty of oscillation modes as well as non-oscillatory components, but only a portion of each is correlated. For positive $\tau$, the correlated portions are between the component traveling from $\bf{r}$ to $\bf{r'}$, $\psi_{\bf r\rightarrow r'}(t)$, and the component at ${\bf r'}$ responding to signals from ${\bf r}$,
$\psi_{\bf r'\leftarrow r}(t)$: 
\begin{equation}
\psi_{\bf r'\leftarrow r}(t)=\psi_{\bf r\rightarrow r'}(t) \circledast G_{|{\bf r}-{\bf r'}|}(t)=\int \psi_{\bf r\rightarrow r'}(t-\tau) G_{|{\bf r}-{\bf r'}|}(\tau) \mathrm{d}\tau, 
\end{equation}
where $G_{|{\bf r}-{\bf r'}|}(t)$ is the Green's function for this distance and $\circledast$ denotes (temporal) convolution. Note that here, $G$ is defined to express the response relative to the component $\psi_{\bf r\rightarrow r'}$, not the whole oscillation $\Psi_{\bf r}$. The uncorrelated components vanish after a large-field average, and we have: 
\begin{equation}
\Phi(|{\bf r}-{\bf r'}|,\tau)=\langle\psi_{{\bf r\rightarrow r'}}(t)\star(\psi_{{\bf r\rightarrow r'}} \circledast G_{|{\bf r}-{\bf r'}|})(t)\rangle_{|{\bf r}-{\bf r'}|}\label{le}.
\end{equation}
In Fourier's domain, this is equivalent to 
\begin{equation}
\mathcal{F}(\Phi_{|{\bf r}-{\bf r'}|})(\nu)=\langle|\mathcal{F}(\psi_{{\bf r\rightarrow r'}})(\nu)|^2\rangle_{|{\bf r}-{\bf r'}|}\cdot\mathcal{F}(G_{|{\bf r}-{\bf r'}|})(\nu),
\end{equation}
where $\mathcal{F}(\cdot)$ represents (temporal) Fourier transform. 
The component $\psi_{\bf r\rightarrow r'}$ contributes a fraction $\alpha$ in power of the whole oscillation $\Psi_{\bf r}$, that is:
\begin{equation}
 \begin{split}
\langle|\mathcal{F}(\psi_{{\bf r\rightarrow r'}})(\nu)|^2\rangle_{|{\bf r}-{\bf r'}|}&=\alpha_{{\bf r\rightarrow r'}, \nu}\langle|\mathcal{F}(\Psi_{\bf r})(\nu)|^2\rangle_{|{\bf r}-{\bf r'}|}=\alpha_{{\bf r\rightarrow r'}, \nu}\mathcal{F}(\Phi_0)(\nu) \\
\alpha_\nu&=\int_0^{\infty} \alpha_{{\bf r\rightarrow r'}, \nu}\  \mathrm{d}{\bf{r'}}^2=1,
\end{split}
\end{equation}
where  $\Phi_0=\Phi(0,\tau)$ is a self-covariance function calculated from Equation~(\ref{cov}). The fraction $\alpha_{{\bf r\rightarrow r'}, \nu}$ describes the power distribution of the component of wave that travels to different locations (corresponding to different modes) for a certain frequency.   The Green's function is then 
 \begin{equation}
 \begin{split}
 &  \alpha_{{\bf r\rightarrow r'}, \nu}\mathcal{F}(G_{|{\bf r}-{\bf r'}|})(\nu)  =\frac{\mathcal{F}(\Phi_{|{\bf r}-{\bf r'}|})(\nu)}{\mathcal{F}(\Phi_0)(\nu)} \\
\mathrm{or,}\qquad\qquad  &G_{|{\bf r}-{\bf r'}|}(t)  = \mathcal{F}^{-1}(\frac{\mathcal{F}(\Phi_{|{\bf r}-{\bf r'}|})(\nu)}{\alpha_{{\bf r\rightarrow r'}, \nu}\mathcal{F}(\Phi_0)(\nu)})(t),
\label{Gf}
\end{split}
 \end{equation}
where $\mathcal{F}^{-1}(\cdot)$ represents inverse Fourier transform.
This way we calculate the three sets of Green's function $G_{\mathrm{umbra}}$, $G_\mathrm{penumbra}$, and $G_\mathrm{quiet}$ for outgoing waves from sunspots or near-sunspot regions by confining ${\bf r}$ in umbra, penumbra, and quiet regions, respectively, with the unknown $\alpha$'s  that will be dealt with in Section~\ref{sec_reconstruct}. The calculation is done between 2~and~7~mHz, outside of which there is not enough oscillatory power to obtain a good signal-to-noise ratio of  $\Phi$.  Figure~\ref{GF_fig} shows the results of $\alpha_{\bf r\rightarrow r'}G_{\bf |r-r'|}$ where $\alpha_{\bf r\rightarrow r'}$ is frequency-independent. 


\subsection{Oscillatory Velocity Reconstruction}
\label{sec_reconstruct}

In helioseismic holography, $G^+$ is used to reverse waves backward in time from receiver to source, and $G^-$ is used to propagate waves forward. When dissipation is negligible, the wave propagation process is time-reversal invariant, i.e., $G^+(t)=G^-(-t)$, as pointed out by \citet{Lin00}. However, this is not the case in sunspots, 
and we therefore use deconvolution with Green's functions to reverse the waves back in time, that is,
 \begin{equation} 
 \begin{gathered}
 \mathcal{F}(\psi_\mathrm{source})=\mathcal{F}(\psi_\mathrm{receiver})\cdot\frac{1}{\mathcal{F}(G)}\\
 \psi_\mathrm{source}=\psi_\mathrm{receiver}\circledast \mathcal{F}^{-1}(\frac{1}{\mathcal{F}(G)})=\psi_\mathrm{receiver}\circledast G',
\end{gathered}
\label{Gp}
 \end{equation}
{where convolving with $G$ is the forwarding waves from a source to a receiver, and deconvolving with $G$, i.e., convolving $G'$ , is reversing waves from the receiver to the source. In our case, forwarding in time is for sunquake waves to propagate outward from a sunquake source in or near a sunspot to the outside regions, so we need the $G$ for outgoing waves propagating from sunspot regions, $G_{out}$, which has been calculated in Section~\ref{sec_Green}. The corresponding $G'_\mathrm{out}$ can be calculated by Equation~(\ref{Gp}), and we can then} use $G'_\mathrm{out}$ to reconstruct the oscillatory velocity $v$ at the sunquake source: 

 \begin{equation} 
 \begin{aligned}
 v({\bf r},t)&=\int \Psi({\bf{r'}},t)\circledast G'_\mathrm{out}(|{\bf r}-{\bf r'}|,t)\ \mathrm{d}{\bf{r'}}^2\\
  \mathcal{F}(v)({\bf r},\nu)
 & =\int  \mathcal{F}(\Psi)({\bf{r'}},\nu)  \frac{1}{\mathcal{F}(G_\mathrm{out})(|{\bf r}-{\bf r'}|,\nu)}\ \mathrm{d}{\bf{r'}}^2 \\
 & =\int \alpha_{{\bf r\rightarrow r'}, \nu}  \cdot \frac{\mathcal{F}(\Psi)({\bf{r'}},\nu)}{\alpha_{{\bf r\rightarrow r'}, \nu}\mathcal{F}(G_{\mathrm{out}|{\bf r}-{\bf r'}|})(\nu)}\ \mathrm{d}{\bf{r'}}^2, 
 \end{aligned}
 \label{v}
 \end{equation}
where the denominator $\alpha_{{\bf r\rightarrow r'}, \nu}\mathcal{F}(G_{\mathrm{out}|{\bf r}-{\bf r'}|})(\nu)$ can be calculated from Equation~(\ref{Gf}). 

Now the only thing left is the unknown $\alpha_{{\bf r\rightarrow r'}, \nu}$, which describes the fraction of oscillation power of frequency $\nu$ at ${\bf r}$ that travels to ${\bf r'}$. The $\alpha_{{\bf r\rightarrow r'}, \nu}$ is expected to decay with the increase of $|{\bf r}-{\bf r'}|$, and within the measurement frequency band, 2-7 mHz band, the frequency dependence of the distribution over $|{\bf r}-{\bf r'}|$ is very weak. Therefore in practice, we choose a frequency-independent weight $w_{|{\bf r}-{\bf r'}|}$ that sums to 1 to substitute $\alpha_{{\bf r\rightarrow r'}, \nu}$, and then integrate over a certain range to obtain:
 \begin{equation} 
 \begin{aligned}
\mathcal{F}(v)({\bf r},\nu)=\int_{a<|{\bf r}-{\bf r'}|<b}& w_{|{\bf r}-{\bf r'}|}  \cdot \frac{\mathcal{F}(\Psi)({\bf{r'}},\nu)}{\alpha_{{\bf r\rightarrow r'}, \nu}\mathcal{F}(G_{\mathrm{out}|{\bf r}-{\bf r'}|})(\nu)}\ \mathrm{d}{\bf{r'}}^2 \\
=\int_{a<|{\bf r}-{\bf r'}|<b}& w_{|{\bf r}-{\bf r'}|} \cdot  \frac{\mathcal{F}(\Psi)({\bf{r'}},\nu) \cdot \mathcal{F}(\Phi)(0,\nu))}{\mathcal{F}(\Phi)({|{\bf r}-{\bf r'}|},\nu)}\mathrm{d}{\bf{r'}}^2 \\
  \end{aligned}
  \label{final}
\end{equation}
\begin{equation}
\int_{a<|{\bf r}-{\bf r'}|<b}w_{|{\bf r}-{\bf r'}|} \ \mathrm{d}{\bf{r'}}^2 =1,
\end{equation}
where $a$ and $b$ define the holography pupil, between 15 Mm and 100 Mm in our case. Note that this integration is not theoretically complete unless the holography pupil covers the whole sphere, which is not practical since the sunquake ripples outside the selected holography pupil is too weak to be detected. Therefore, the oscillations reconstructed are limited to only high-$l$ helioseismic oscillation modes.  As $|{\bf r}-{\bf r'}|$ increases, the $\Phi(|{\bf r}-{\bf r'}|,\tau)$ in the denominator of Equation~(\ref{final}) is noisier and intrinsically smaller in value, thus a proper weight is necessary to ensure the result does not explode. 
The $w_{|{\bf r}-{\bf r'}|}$ we use is determined by the signal-to-noise ratio of $\Phi(|{\bf r}-{\bf r'}|,\tau)$,  which decays roughly with a power of 2 with $|{\bf r}-{\bf r'}|$, and is also a reasonable approximation of the expected decay of $\alpha_{\bf r\rightarrow r'}$ with $|{\bf r}-{\bf r'}|$. 
{We acknowledge that, however, our reconstructed velocity might be subject to a scaling factor due to the imperfect approximation of $\alpha_{\bf r\rightarrow r'}$'s.}

\section{Data}
\label{sec_Data}

For this study, {to obtain a sample size of $\sim$30 sunquake events,} we use  {the} HMI observations to survey 60 strongest flares{, which }occurred within 73\degr of longitude, were stronger than M6.3 in X-ray flux, and had a clear intensity enhancement in the HMI line-core intensity data. { Such a selection of longitude ensures spatial resolution of the vicinity of the flaring sites is still resolvable, and our line-of-sight observations can still be used to calculate the sunquake energy.}

These include 36 X-class flares and 24 M-class flares. 
For each event, we track the line-of-sight magnetic field ($B$), continuum intensity ($Ic$), 
line-core intensity ($LC$), and Doppler velocity ($V$) data with the Carrington rotation rate, 
and project the images into Postel's coordinates with the main sunspot at the field center. 
All the tracked maps have a size of 1024$\times$1024 pixels with a spatial resolution of $0\fdg03$ 
($\sim$0.36 Mm) pixel$^{-1}$ and a duration of 8 hours with a temporal cadence of 45 s. 

\begin{figure}[!t]
\epsscale{0.99}
\plotone{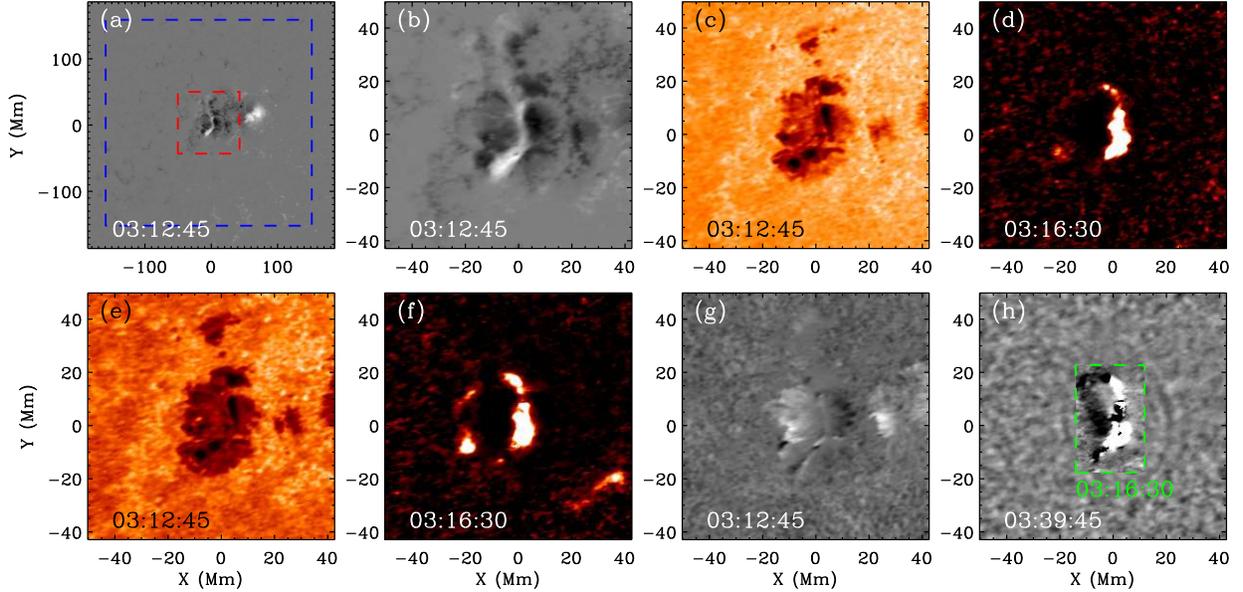}
\caption{Examples of observation data for an X1.8 flare at 03:15:45 UT of 2012 October 23, centered 
at S13E58 on the solar disk. The images are (a) magnetic field,  (b) magnetic field in the red 
box in panel (a), (c) continuum intensity, (d) continuum intensity running difference, (e) line-core 
intensity, (f) line-core intensity running difference, (g) Doppler velocity, and (h) Doppler velocity 
running difference {inside the green box, and masked and filtered Doppler velocity data outside the green box.}}
\label{map}
\end{figure}

Figure~\ref{map} shows selected data snapshots for an X1.8 flare of 2012 October 23, which serves as 
an example for the analyses in this paper. Figure~\ref{map}a shows the line-of-sight magnetic field for 
the whole field, with the red box indicating the sunspot region where the oscillatory velocities 
will be reconstructed in Section~\ref{sec_snap} using the data in the blue box.  
All the rest panels are zoomed in to show 
only the red-box region. Raw data of $B$, $Ic$, $LC$, and $V$ at a few minutes before the 
flare onset are shown in Figures~\ref{map}a-\ref{map}c, \ref{map}e and \ref{map}g. During the flaring time, there were 
significant changes in $Ic$, $LC$ and $V$, as shown in the corresponding running-difference 
images (image difference between consecutive time steps) in Figures~\ref{map}d, \ref{map}f and \ref{map}h {(inside green box)}.  The continuum 
intensity enhancement in Figure~\ref{map}d is used to detect HMI continuum emission, as shown in this example; 
the line-core intensity enhancement in Figure~\ref{map}f is used for spatial and temporal signatures of 
the flare; and the Doppler transients in Figure~\ref{map}h {(inside green box)}, displayed with extreme values, will 
be compared with the reconstructed sunquakes for their spatial relation. The Doppler-velocity 
data are further processed before being fed into our holography calculation: the Doppler 
transients are masked (substituted by mean background values) to prevent possible contamination to the sunquake-signal reconstruction; and a filter is applied to remove convection, which is irrelevant to the sunquake waves and not needed in the sunquake reconstruction. 
{The roles of masking Doppler transients are, first, to prevent the leakage of Doppler transient values into their surrounding area when applying a spatial and temporal filtering in the Fourier domain; and second, to avoid deconvolving the extreme values in the Doppler transients with Green’s functions, which can produce ring-shaped artifacts in the reconstructed wave-field.}
{After masking and filtering, the sunquake ripples are found to be clearer in the processed Doppler velocity data than in the raw or running-difference Doppler velocity data. We thus also show the ripple of the example event in Figure~\ref{map}h (outside green box) at about 23 min after the flaring time. }
{Note that, in this study, we use these masked and filtered Doppler velocity data  (without running difference) for reconstructing the oscillatory velocity and other calculations, and use the running-difference Doppler velocity data (without masking and filtering) only for visualizing the Doppler transients. }

\section{Results: An Example Study}
\label{sec_res}

Oscillatory velocity within $\pm1$ hour around the flaring time in the sunspot area,  as denoted by red box  in Figure~\ref{map}a, are reconstructed using 3.2-hour-long (masked and filtered) Doppler-velocity observations within the blue box. 
{The 3.2-hour long data is so chosen that it includes the $\pm1$ hour around the flare and the 75-min-long Green’s function.} 
{Since the Doppler velocities are along the line of sight and the flaring sites at various locations on the solar disk have different line-of-sight projections, we further scale the reconstructed velocity by cosine of the disk-centric distances (in unit of spherical angles) to correct the projection effect. This gives proxies of radial velocity, assuming that the oscillations are intrinsically dominated in the radial direction.}
{For further studies,} the reconstructed {radial velocities} are then separated into two frequency bands: 3-5 mHz and 5-7 mHz, {the former of which is usually dominated by background oscillations and in the latter, sunquake signals more often stand out from the relatively weaker background signals.}


\begin{figure}[!t]
\epsscale{0.90}
\plotone{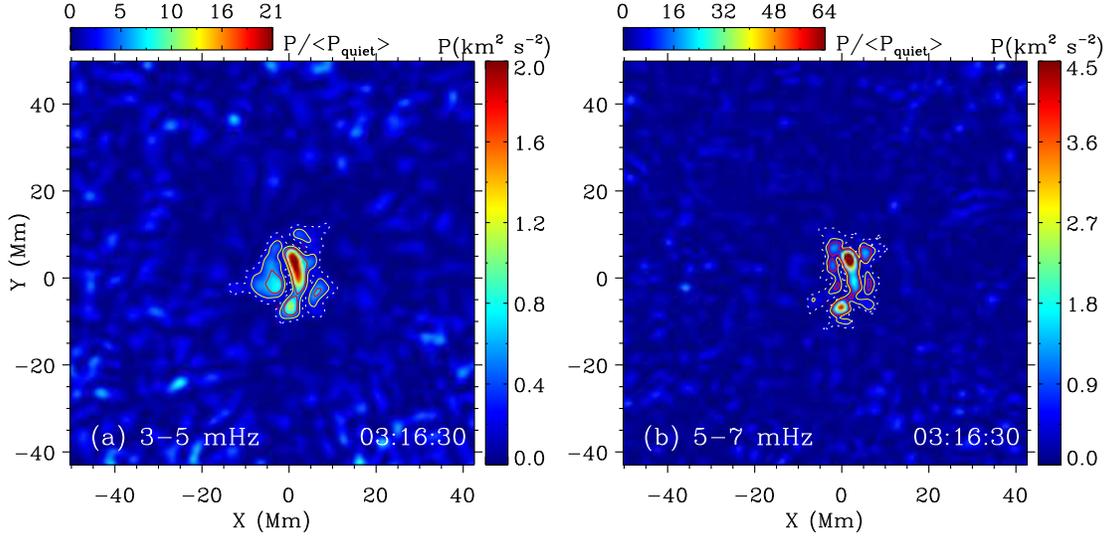} 
\caption{Egression power maps of the reconstructed oscillatory velocities for the X1.8 flare on 2012 October 23, in (a) $3-5$ mHz and (b) $5-7$ mHz. Vertical colorbars show absolute values of the power, and horizontal colorbars show the ratios to the quiet-Sun mean values. Red and yellow contours show the 3$\sigma_\mathrm{qt}$ and 3$\sigma_\mathrm{ar}$ levels in each frequency band, respectively, inside the sunquake area enclosed by white dashed contours at 1$\sigma_\mathrm{qt}$ level.}
\label{pow_fig}
\end{figure} 

\begin{figure}[!t]
\epsscale{0.90}
\plotone{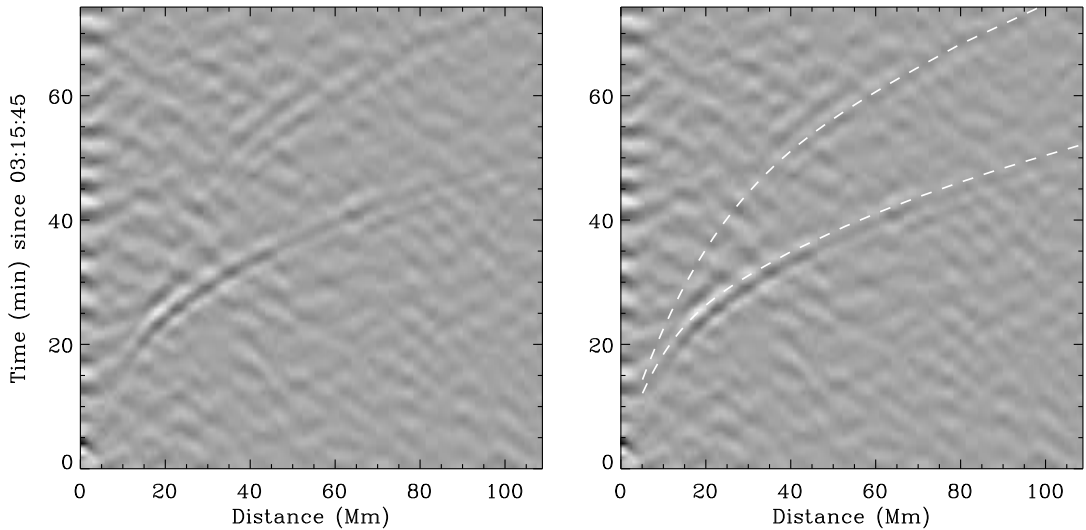} 
\caption{Left: Space--time diagram of the sunquake wave excited by the X1.8 flare on 2012 Oct 23. Right: Same as left panel, but with white dashed lines indicating the theoretical distance-dependent first- and second-skip travel times for acoustic waves.}
\label{td}
\end{figure}

\subsection{Sunquake Detection}
\label{sec_pow}

A sunquake is recognized by a certain criterion on the egression power $P$, defined as 
\begin{equation}
P(t)=\frac{2}{W}\int_{-\frac{W}{2}}^{\frac{W}{2}}v^2(t+\tau)\ \mathrm{d}\tau,
\end{equation}
where $W$ is an averaging window and $v(t)$ is the reconstructed oscillatory velocity. We choose $W$ to be five time steps, i.e., 225 s, slightly larger than the period of $v^2$ (half period of $v$) for low-frequency oscillations, and a normalization factor of 2, so that square root of $P$ can be approximated to be the oscillation amplitude (envelope of a wave packet). second-skip
Figure~\ref{pow_fig} shows power maps for two frequency bands at the peak-power time, each displayed in two color scales: by absolute values of $P$ and by ratio of $P$ to the quiet-Sun mean power. In this case the maximum {reconstructed (radial)} velocity is over 1~km s$^{-1}$ in both frequencies, and the maximum power is about 25 and 92 times the quiet-Sun mean for the two frequency bands, respectively. Statistics of the background values between 60 min to 15 min before the flaring time, in both quiet regions and in sunspots, are calculated as references to compare with the egression power of the sunquake source.  The red and yellow contours in Figure~\ref{pow_fig} indicate the 3$\sigma$ (99.7 percentile) of $P$ in the quiet-Sun (3$\sigma_\mathrm{qt}$) and in the active region (3$\sigma_\mathrm{ar}$) in each frequency band, respectively, which are about  6.4 and 3.1 times of the quiet-Sun mean in $3-5$ mHz, and 10.3 and 6.2 times of the quiet-Sun mean in $5-7$ mHz in this case. We call the areas enclosed by the contours 3$\sigma_\mathrm{qt}$ kernels  and 3$\sigma_\mathrm{ar}$ kernels, respectively. A flare is labeled as sunquake active if there are 3$\sigma_\mathrm{qt}$  kernels with {sizes $\geq 25$ pixels} close to or coinciding with a flare ribbon during the flaring time, in at least one of the two frequency bands. {If the 3$\sigma_\mathrm{qt}$ kernels are smaller, or if the peak power does not exceed 3$\sigma_\mathrm{qt}$ but exceeds 3$\sigma_\mathrm{ar}$, the acoustic source is considered as a weak source. For all weak sources,} we further examine their space--time diagrams, like the one shown in Figure~\ref{td}, which is the time sequences averaged from annuli and stacked in distances, and search for visible sunquake signals to determine whether a sunquake occurs. All the sunquakes that are detected {in this study} are confirmed by space--time diagrams. 
{Note that, some sunquake candidates are found to have both ``weak" acoustic sources, and suspicious sunquake signals on the space--time diagrams. They could otherwise be true sunquakes, but we have excluded them in our study to obtain a robust sunquake sample to verify our hypothesis. }
Figure~\ref{td} shows the space--time diagram of this event, in which the sunquake waves, even the second-skip waves, are clearly visible, with the travel time consistent with a theoretical ray-path approximation \citep{Dsi96}. 
{The second-skip wave corresponds to the sunquake waves that get reflected by the surface when first approaching the surface from the interior, and propagate one more skip through the solar interior before reaching the surface again. The second-skip sunquake waves can only be observed in a few very strong sunquakes. }

To estimate the sunquake egression energy, we integrate the acoustic energy flux $\frac{1}{2}\rho c_{\rm s} v^2$ in an area encompassing the kernels, determined by 1$\sigma$ (68.3 percentile) of quiet-region background and delimited by white dashed contours in Figure~\ref{pow_fig}, and over a 30-min window around sunquake power peak time, an approach similar to  \citet{Zkv13}.  The 30-min window is so chosen to cover the entire wave train of the reconstructed signal (see Section~\ref{sec_curve} and Figure~\ref{curves_fig}). We use the density $\rho$ and sound speed $c_\mathrm{s}$ in Model S \citep{Chr96} at the height of 250 km. The estimated total energy for this sunquake is $1.4\times10^{29}$ erg, which is among the most powerful sunquake events in Solar Cycle 24 (see Table 1). Note that the energy estimate depends sharply on the area and time chosen for integration{, and could also be influenced by the scaling factor due to the calibration uncertainty of Green’s functions that we acknowledged in Section 2}.  Our energy estimate for this event is comparable to that reported by \citet{Sha17}, and the estimates for the sunquakes in the 2011 February 15 flare (see Table 1) are compatible to those estimated by \citet{Zkv13}, indicating that our derived Green's function is reasonable in oscillatory amplitudes. Note that our energy estimates are expected to be larger than those estimated by the holography method with theoretical-calculated Green's functions, because the observation-based Green's functions include dissipation naturally. Dissipation of acoustic waves eliminates the sunquake energy that arrives at the surface where sunquake ripples are observed, and the corresponding energy loss could be a non-trival fraction in sunspot regions.

\begin{figure}[!t]
\epsscale{0.99}
\plotone{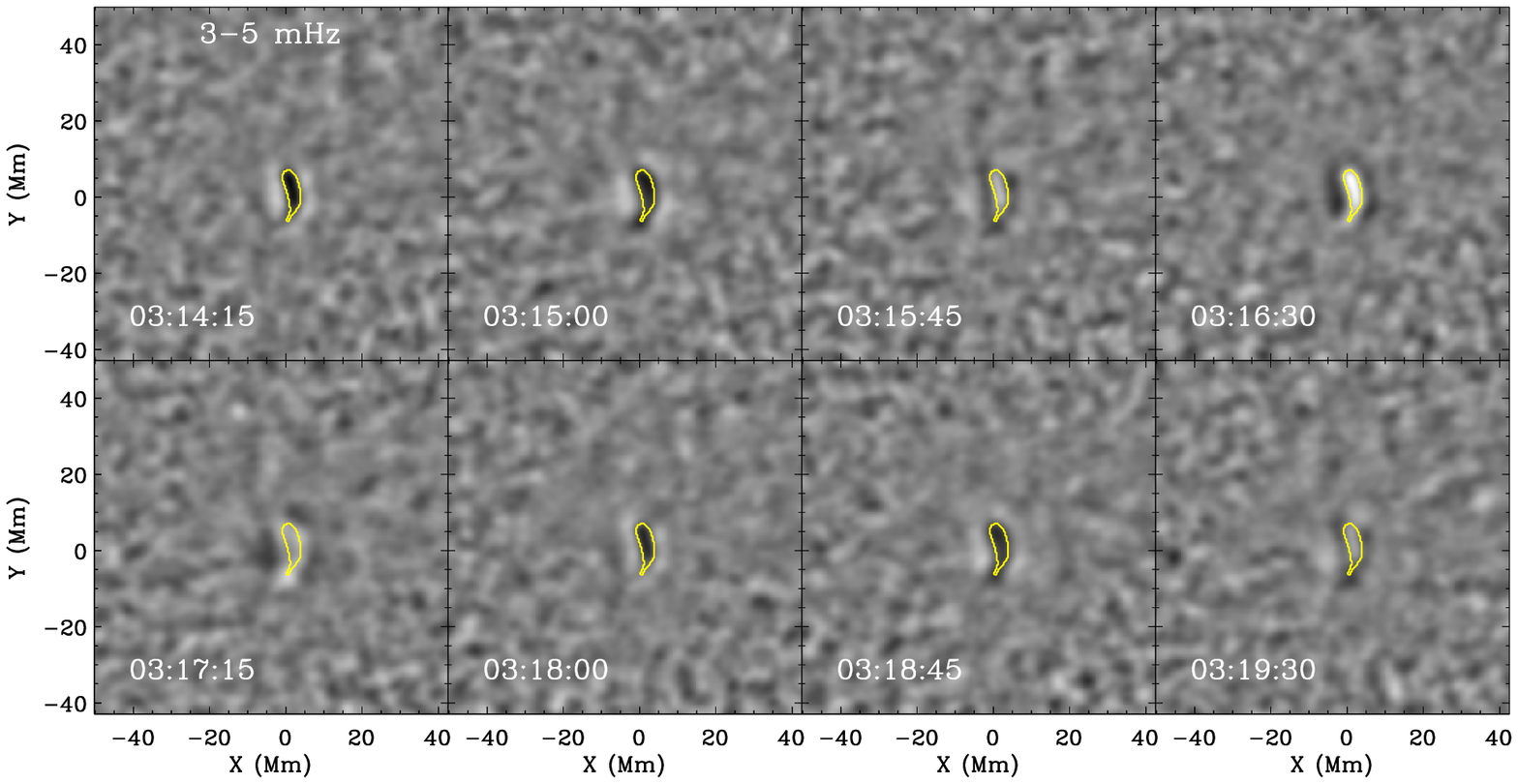}
\plotone{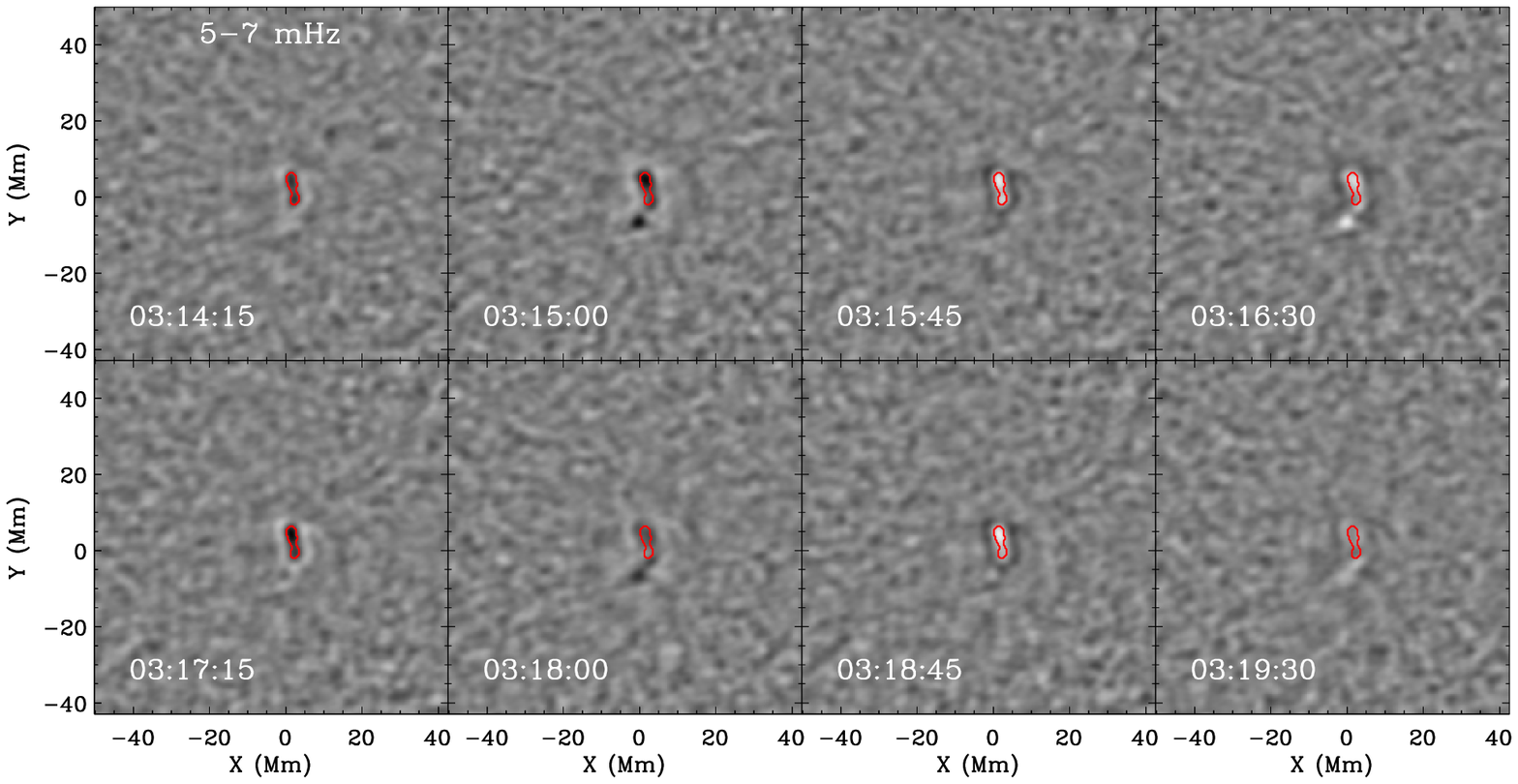}
\caption{Snapshots of the reconstructed oscillatory velocity maps in $3-5$ mHz (top) and $5-7$ mHz  (bottom) during the flaring time. The $3-5$ mHz and $5-7$ mHz oscillation cores are plotted in yellow and red, respectively.
Positive (white) values represent downward velocity, and negative (dark) values represent upward velocity. The velicty is displayed with a range of $-500$ m s$^{-1}$ to $500$ m s$^{-1}$. }
\label{snap_fig}
\end{figure} 

\begin{figure}[!t]
\epsscale{0.99}
\plotone{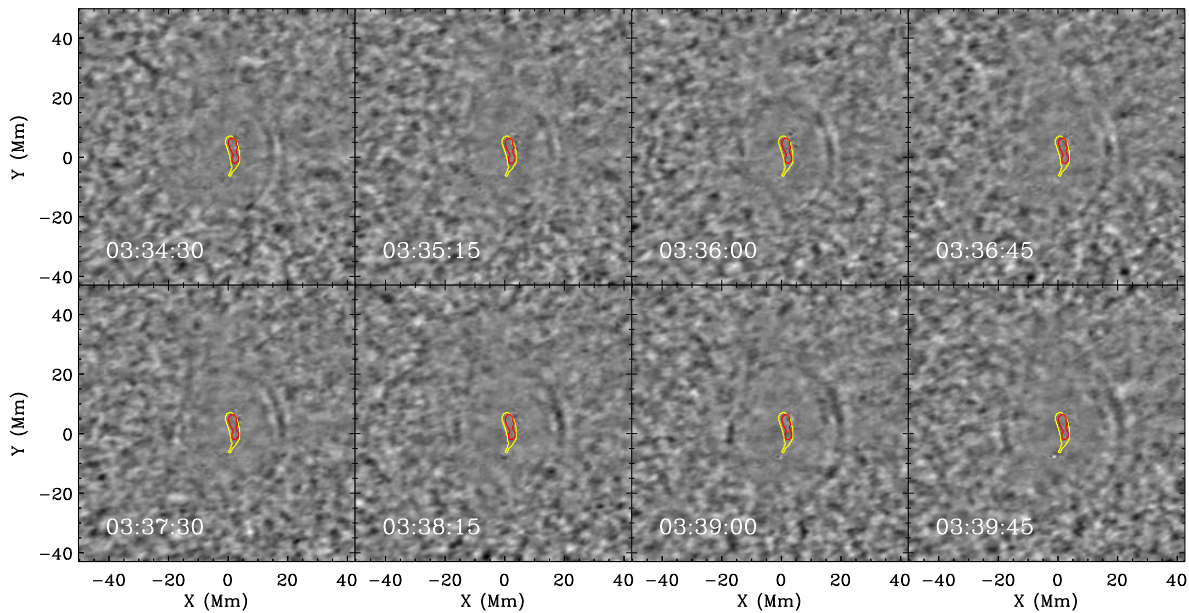}
\caption{Snapshots of the masked and filtered Doppler velocity images. The $3-5$ mHz and $5-7$ mHz oscillation cores are plotted in yellow and red, respectively. These snapshots are taken about 20 min after the flare onset when the sunquake ripples become visible.}
\label{snap_ripple}
\end{figure} 

\begin{figure}[!t]
\epsscale{0.99}
\plotone{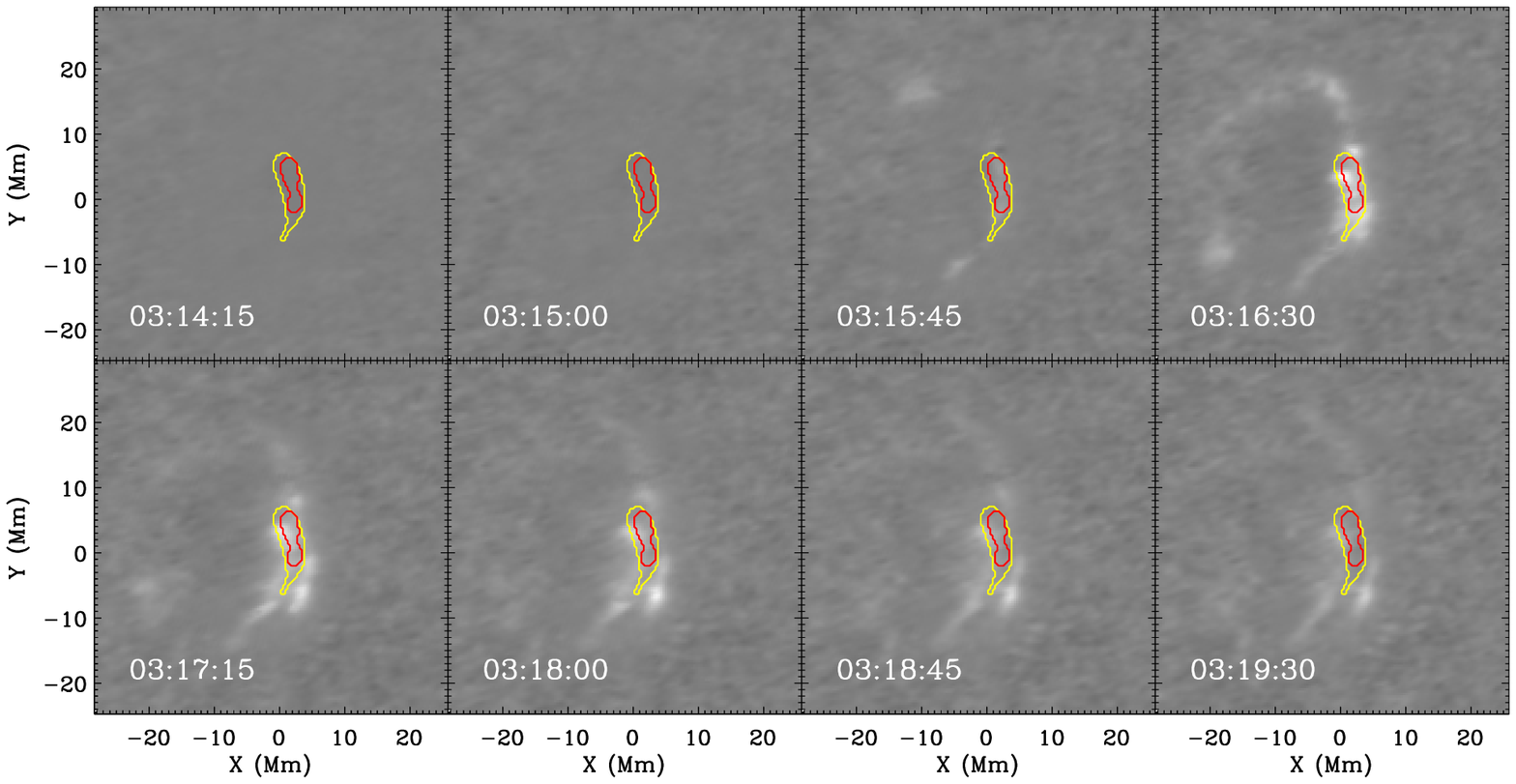}
\plotone{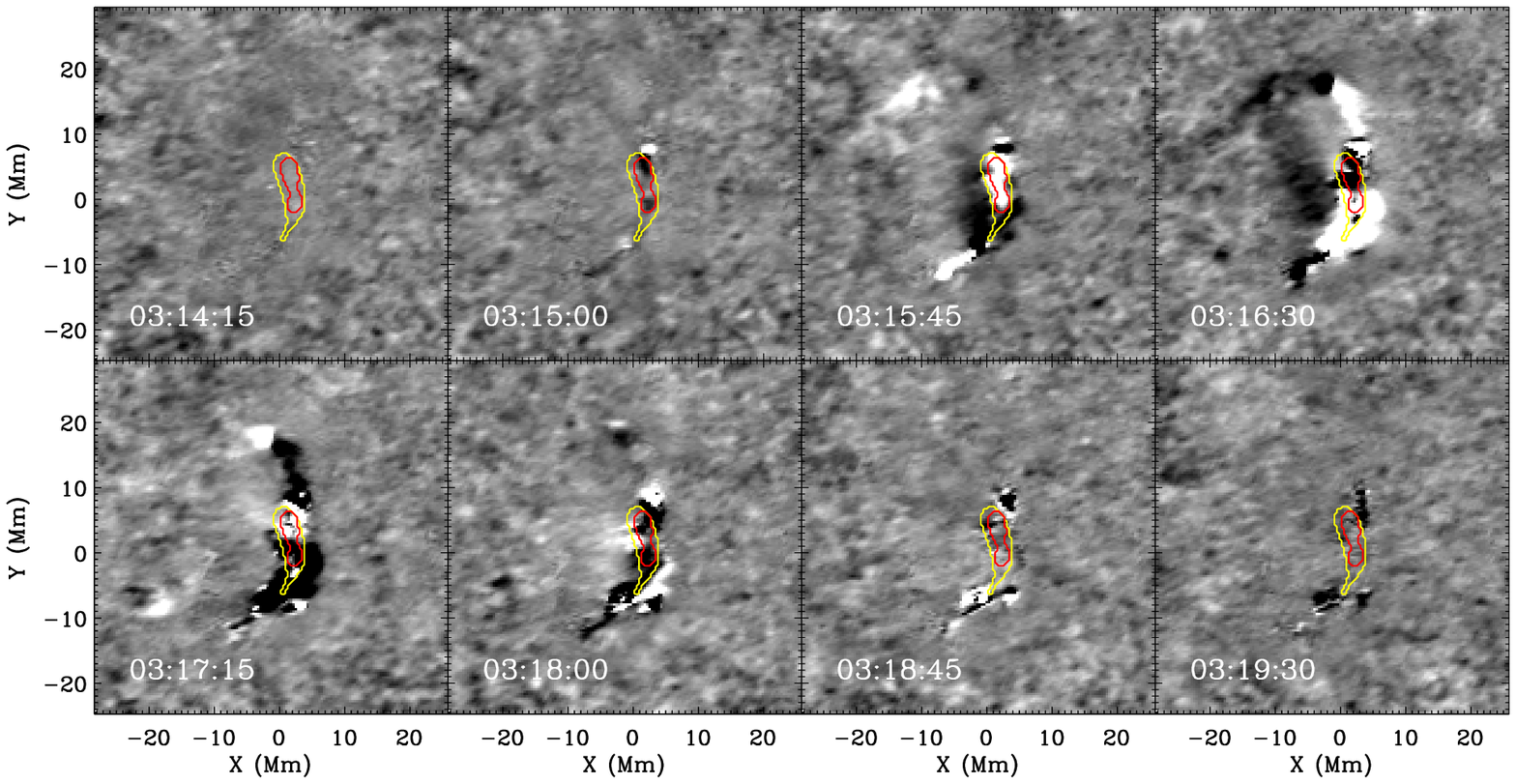}
\caption{Snapshots of the line-core intensity with background subtracted (two top rows) and Doppler velocity running-difference images (two bottom rows). The $3-5$ mHz and $5-7$ mHz oscillation cores are plotted in yellow and red, respectively. These snapshots are taken near the flare peak time. 
{The field of view is smaller than that in Figures~\ref{pow_fig}, \ref{snap_fig}, \ref{snap_ripple}, \ref{all_M}, and \ref{all_Ic} to zoom in the flare sites for a better spatial comparison.}}
\label{snap_fig2}
\end{figure} 


\subsection{Oscillatory Velocity Field at the Sunquake Source}
\label{sec_snap}
Next, we study the temporal and spatial relation between the flare and the reconstructed oscillatory velocity at the detected sunquake source. The reconstructed velocity maps of the example sunquake during the flaring time and in both frequency bands are shown in Figure~\ref{snap_fig}.
A distinct sunquake source is detected oscillating upward and downward in both frequencies. To examine our hypothesis introduced in Section~\ref{sec1}, we  integrate the sunquake velocity during the flare's impulsive phase and examine its oscillatory direction. However, in extended sunquake kernels like this example, sometimes the central sunquake area is surrounded by a halo of oscillations with an opposite phase, complicating the integration of velocity and power. The halos are believed to be an artifact unavoidably caused by the following two scenarios. First, at a given moment, the photospheric response to a $\delta$-function-like impulse is wide spread, containing a few ripples. Deconvolving these ripples with a Green's function, truncated in both frequency and wavenumbers, to reconstruct the wave source will result in a spread source surrounded by oscillations with opposite phases. Second, many sunquake ripples are non-circular, influenced by the shape of flares and sunspot geometry case by case. These non-circular ripples will enhance the side-lobes forming around the source. Therefore, integrating velocity across the whole sunquake kernel may give a misleading net value. 
In practice, we select smaller areas, where the egression power peaks and whose space--time diagram is clearest, as oscillation ``cores" for integration of velocity. Examples of such core areas are shown in the contoured area in Figure~\ref{snap_fig}. For some events in which the egression is complicated by irregular or complex sunquake ripples, the determination of oscillation cores also includes a careful examination on the location of sunquake ripples. 

Figure~\ref{snap_ripple} shows the sunquake ripples of the example event about 20 min after the flare onset,  with the oscillation cores of the two frequency bands plotted as contours. The ripples become visible starting from about 15 Mm away from the sunquake source, and are visible on both east and west sides but stronger on the west (closer to the disk center). The ripples also form an elliptical shape, indicating a bar-shape sunquake source,  consistent with the contoured oscillation core area.

\begin{figure}[!t]
\epsscale{1}
\plotone{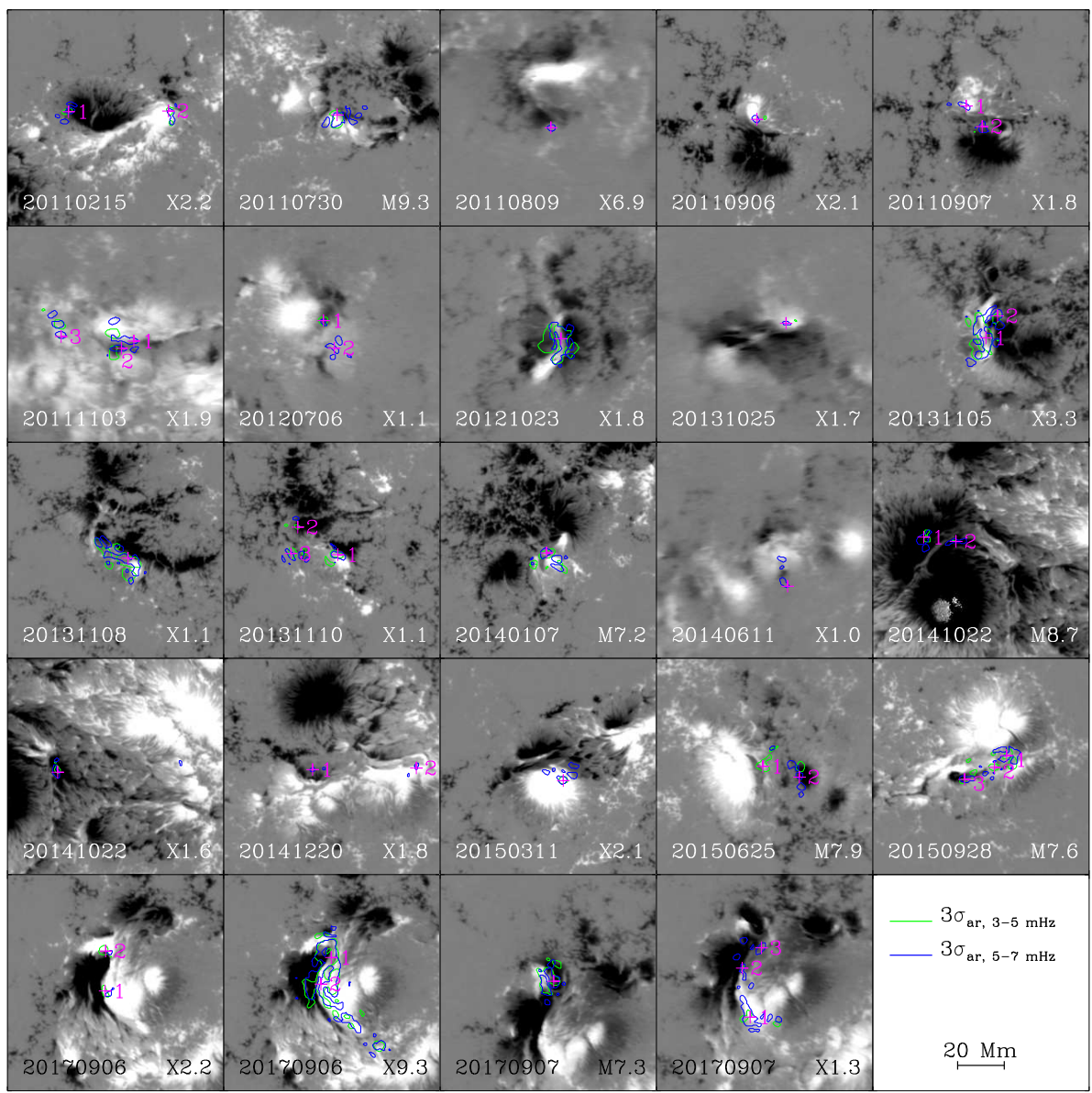}
\caption{Line-of-sight magnetic field maps of all the 24 active regions that host sunquake-generating flares. Green and blue contours show the 3$\sigma_\mathrm{ar}$ levels of the sunquake egression power in $3-5$ mHz and $5-7$ mHz, respectively. The `+' marks denote the centroids of oscillation cores for each sunquake. A number is marked beside the `+' sign when more than one sunquake source is identified in the active region.}
\label{all_M}
\end{figure}

Figure~\ref{snap_fig2} shows the line-core intensity snapshots with background subtracted, as well as the Doppler running difference images, in the same time frames as Figure~\ref{snap_fig} for comparisons. The flare starts to show $LC$ enhancement at 03:15:45 UT and peaks in $LC$ at 03:16:30 UT, during which times the velocities in oscillation cores in both frequencies (Figure~\ref{snap_fig}) are downward, consistent with our hypothesis in Section~1. As for the spatial relation with the photospheric response to the flare, this sunquake occurs in the area of maximum $LC$ enhancement {(top panels of Figure~\ref{snap_fig2})} and on the polarity inversion lines (PILs, see Figures~\ref{map} and \ref{all_M}). Spatial relation of more sunquakes and flares will be discussed in Section~5. On the Doppler running-difference maps {(bottom panels of Figure~\ref{snap_fig2})}, severe anomalies occur throughout the flaring period and overlap with the flare ribbons in a large area; but the sunquake source only overlaps with part of the Doppler transients. 


\subsection{Temporal Comparison}
\label{sec_curve}

\begin{figure}[!t]
\epsscale{0.9}
\plotone{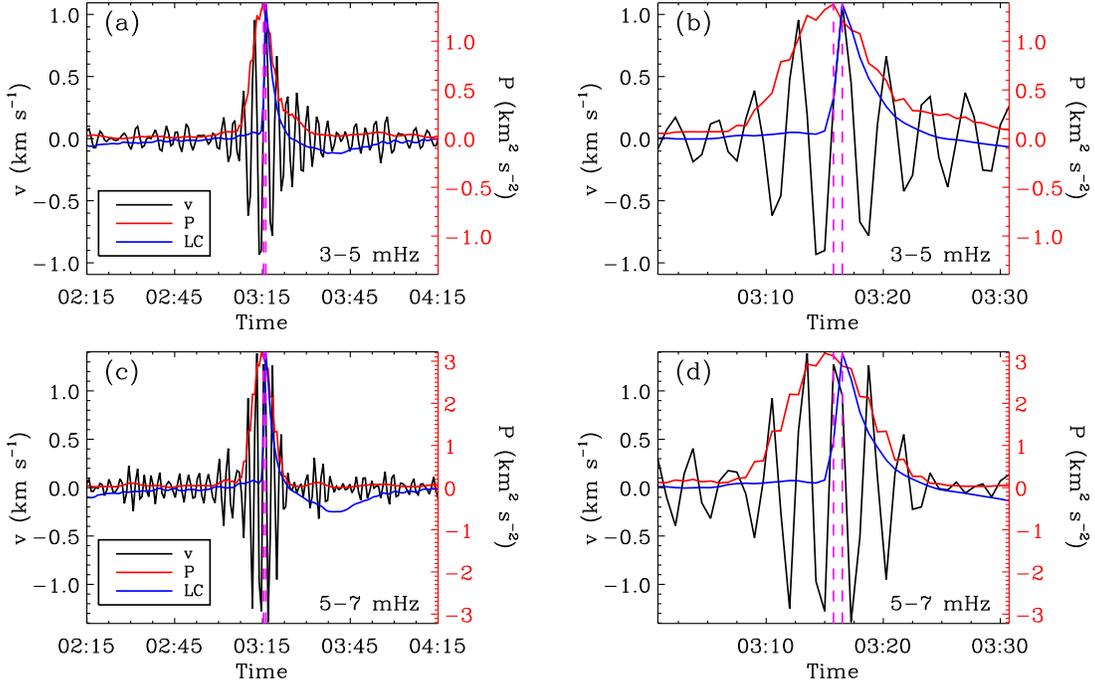}
\caption{Temporal evolutions of the oscillatory velocity, oscillatory power $P$, and line-core intensity integrated in the oscillation cores for (a) $3-5$ mHz, (b) $3-5$ mHz zoomed in, (c) $5-7$ mHz, and (d) $5-7$ mHz zoomed in. Between magenta dashed lines in each panel is the flare's impulsive phase determined for the oscillation cores. }
\label{curves_fig}
\end{figure} 

To examine the temporal relation between the sunquake and the flare, we compare the temporal evolutions of the oscillatory velocities, oscillatory power, and  line-core intensity, all of which are obtained from averaging these quantities inside the oscillation core. In Figure~\ref{curves_fig}, the velocity and power curves are plotted in black and red, respectively, and the $LC$ curve is plotted after the removal of the background and normalization. The power curve has a FWHM of about 8 min, which is due to the finite bandwidth, $\Delta t=\frac{1}{\Delta \nu }=\frac{1}{2\ \mathrm{mHz} }=$~500~s. If plotted in the full frequency band, 2-7mHz, the FWHM will be narrower, but still finite in the absence of low frequency components below 2 mHz.  It is shown that significant oscillation signals are recovered around the $LC$ peak time, however, the sunquake initial impulse is not resolved, limited by the resolution $\Delta t$. But we can still examine the direction of the sunquake {(radial)} velocity. It can be verified that, although the bandpass filter widens the impulse signal, the sign of the signal during the impulsive period is reserved.  Therefore, we examine the sign of the reconstructed oscillatory velocity during the flare's impulsive phase. 

Here we define the flare's impulsive phase as the time when the locally averaged $LC$ curve is between $\frac{1}{e}$ and $1$ time of its peak value. This definition corresponds to the $LC$ brightening in a local area, rather than the X-ray flux or the light intensity of the whole flare. During the impulsive phase of the example flare, shown as between magenta lines in Figure~\ref{curves_fig}, the oscillatory velocities for both frequencies are positive (downward), consistent with our hypothesis of that for a detectable sunquake, flares occur during a downward background oscillatory motion. The average velocities are 1082 m s$^{-1}$ in $3-5$ mHz and 950 m s$^{-1}$ in $5-7$ mHz during the flare's impulsive phase. For the statistical study, the average velocities of all other detected sunquakes are calculated in the same way and summarized in Section~\ref{sec_statistics}.


\begin{figure}[!t]
\epsscale{1}
\plotone{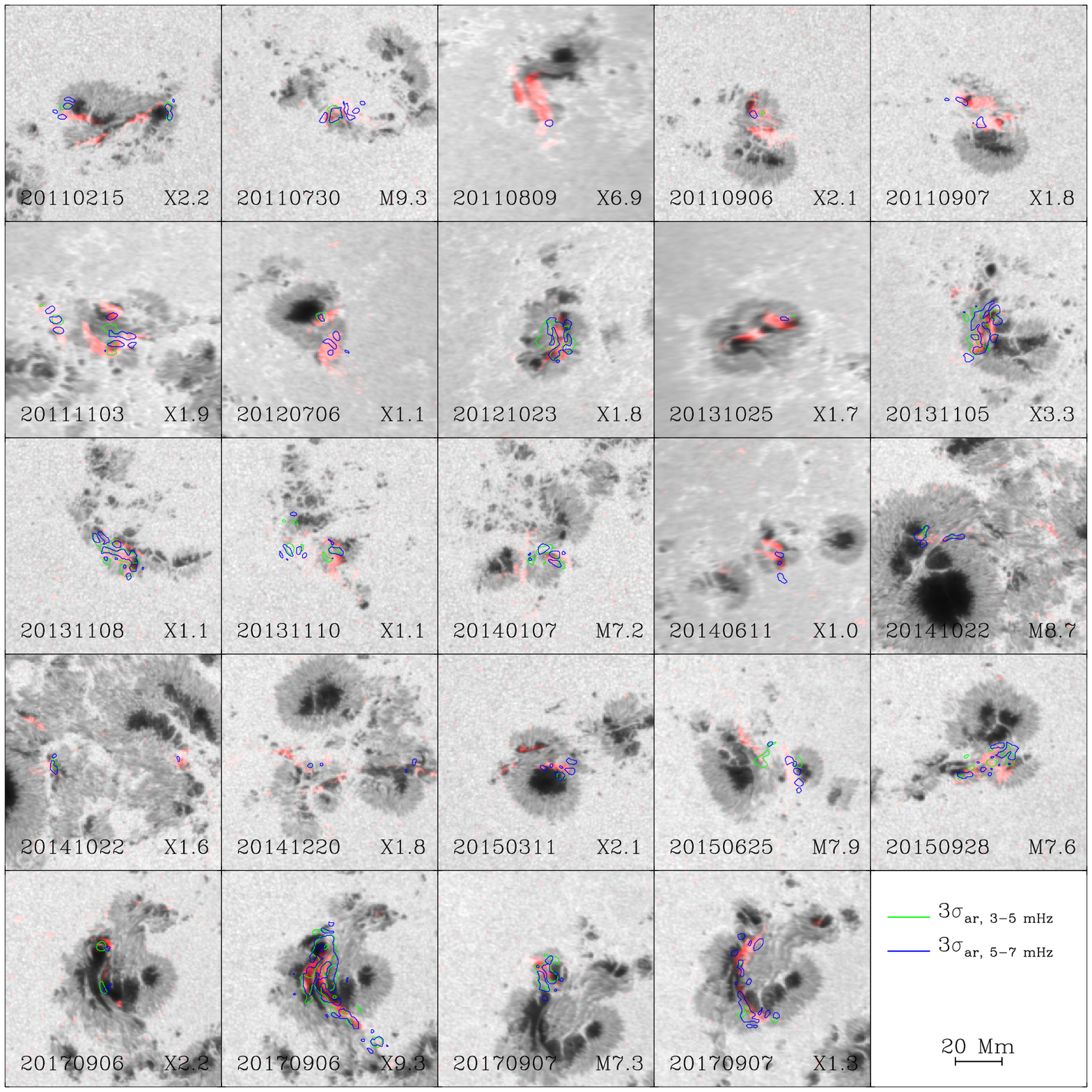}
\caption{Continuum-intensity images of the 24 active regions that host the sunquake-generating flares. The line-core intensity flux are overplotted in red to highlight the locations of the flare ribbons. Green and blue contours show the 3$\sigma_\mathrm{ar}$ levels of sunquake egression power in $3-5$ mHz and $5-7$ mHz, respectively. }
\label{all_Ic}
\end{figure} 

\section{Results: Statistics of All Surveyed Sunquakes}
\label{sec_statistics}

A total of 24 flares, out of 60 flares surveyed, are found to be helioseismic active, giving a total of 41 sunquakes.  Among the 41 sunquakes, 28 of them are detected in both the low and high frequency bands, 10 of them are only active in the high frequency bands, and 3 of them are only active in the low frequency bands. The 3$\sigma_\mathrm{ar}$ egression power kernels of these flares are plotted overlapping their magnetograms (Figure~\ref{all_M}) and continuum intensity images (Figure~\ref{all_Ic}). For flares whose reconstructed oscillatory power is not compact,  multiple sunquakes are counted if two or more kernels, clearly apart from each other, are associated with different flare ribbons or correspond to different (visible) sunquake ripples. The locations of sunquake sources (centroids of oscillation cores) are denoted by `+' marks in Figure ~\ref{all_M}.  A majority of sunquakes occur on or close to PILs.  The locations of sunquake sources are also compared with flare ribbons, denoted by the line-core-intensity flux integrated during the flaring period and depicted in red in Figure~\ref{all_Ic}. Sunquakes tend to occur between or at the edge of flare ribbons. In sunquakes with extended and strong kernels, the kernels may largely overlap with the flare ribbons, but the peak-power locations still tend to be between or at the edge of flare ribbons{, probably indicating the location of the strongest dynamic impacts from the upper atmosphere}. 

Table 1 shows the statistical information of all the flares that are detected sunquake active. The listed longitude (relative to the central meridian) and latitude are for the image centers in Figures~\ref{all_M} and \ref{all_Ic}, and the listed time is for the maximum HMI $LC$ enhancement. Total energies of sunquakes range between $10^{27} -10^{29}$~erg. The last two columns in the table list the average velocities of the oscillation cores of the sunquakes during the flares' impulsive phases,  and the `*' mark following some velocities indicates that the peak egression power of the corresponding sunquake is over 3$\sigma_\mathrm{ar}$ but not over 3$\sigma_\mathrm{qt}$ in that frequency. No velocity is calculated if the peak egression power {in the oscillation core} is less than 3$\sigma_\mathrm{ar}$. It is found that in the $3-5$ mHz frequency band, 25 out of 31 (80.6\%) sunquakes have net downward velocities; in the $5-7$ mHz frequency band, 33 out of 38 (86.8\%) sunquakes have net downward velocities. These roughly agree with the hypothesis proposed in Section 1. Note that in a few events the magnitude of the average velocity is small ($\le$100 m~s$^{-1}$), either because the flare happens to have an impulsive phase longer than one oscillation period, or because the impulsive phase somehow coincides with the time when the oscillatory velocity is close to 0. These data points are not robust. If we exclude them, then 20 out of 25 (80.0\%) sunquakes have net downward velocities in $3-5$ mHz, and 26 out of 28 (92.8\%) sunquakes have net downward velocities in $5-7$ mHz. 
{Note that the difference in percentage (of downward cases) results from the fact that some sunquakes are only counted in one frequency band. Actually, the directions of velocities in two frequency bands are consistent, if we only classify the velocities to be downward or upward when their magnitudes are larger than 100 m~s$^{-1}$, and consider the rest cases to be uncertain (either with a speed less than 100 m~s$^{-1}$, or with no confirmed sunquake).  We then can find that, if a sunquake’s velocity in one frequency band is downward(upward), its velocity in the other frequency band is either also downward(upward), or uncertain.  And overall, out of the 41 detected sunquakes, 29 are downward, 6 are upward, and 6 are uncertain.}

\begin{table}[]
\caption{Statistics of all the sunquakes found in this survey and the flares that triggered them.}
\scriptsize
\centering
\begin{tabular}{c|c|c|c|c|c|c|c|c|c}
\hline
Flare \# & Date     & Class      & AR \# & Longitude & Latitude & Time  & Engergy & v (m~s$^{-1}$) & v (m~s$^{-1}$) \\
         &(yyyymmdd) &            &            &  (deg)& (deg)  & (UT)   & ($\times10^{27}$erg)   &$3-5$ mHz & $5-7$ mHz \\
\hline
1        & 20110215 & X2.2       & 11158 & 12.5      & -19.5    & 01:51:45 & 20.3    &         &         \\
         &          & sunquake 1 &       &           &          &          & 11.7    & 27*     & 29      \\
         &          & sunquake 2 &       &           &          &          & 8.6     & -100*   & -136    \\
2        & 20110730 & M9.3       & 11261 & -34.5     & 15.1     & 02:09:00 & 39.0    & 417     & 337     \\
3        & 20110809 & X6.9       & 11263 & 69.9      & 15.8     & 08:03:00 & 7.4     & /       & -3    \\
4        & 20110906 & X2.1       & 11283 & 18.0      & 14.6     & 22:19:30 & 6.5     & 326*    & 596     \\
5        & 20110907 & X1.8       & 11283 & 31.0      & 14.3     & 22:37:30 & 15.6    &         &         \\
         &          & sunquake 1 &       &           &          &          & 6.8     & /       & 270     \\
         &          & sunquake 2 &       &           &          &          & 8.8     & /       & 111     \\
6        & 20111103 & X1.9       & 11339 & -63.1     & 21.0     & 20:21:45 & 72.7    &         &         \\
         &          & sunquake 1 &       &           &          &          & 28.7    & /       & 193     \\
         &          & sunquake 2 &       &           &          &          & 17.8    & 285     & 311     \\
         &          & sunquake 3 &       &           &          &          & 26.1    & 53    & 177     \\
7        & 20120706 & X1.1       & 11515 & 59.5      & -13.1     & 23:06:45 & 33.5    &         &         \\
         &          & sunquake 1 &       &           &          &          & 10.3    & 387*    & 156     \\
         &          & sunquake 2 &       &           &          &          & 23.2    & /       & 565     \\
8        & 20121023 & X1.8       & 11598 & -58.3     & -12.7    & 03:16:30 & 135.9   & 1082    & 950     \\
9        & 20131025 & X1.7       & 11882 & -72.4     & -8.0    & 07:59:15 & 12.8    & /       & 135     \\
10       & 20131105 & X3.3       & 11890 & -44.1     & -12.6    & 22:12:00 & 93.2    &         &         \\
         &          & sunquake 1 &       &           &          &          & 77.8    & 839     & 775     \\
         &          & sunquake 2 &       &           &          &          & 15.4    & 461     & 1098    \\
11       & 20131108 & X1.1       & 11890 & -14.2     & -13    & 04:25:30 & 69.7    & 344     & 394     \\
12       & 20131110 & X1.1       & 11890 & 13.6      & -13.1    & 05:13:30 & 50.8    &         &         \\
         &          & sunquake 1 &       &           &          &          & 22.8    & 445     & 508     \\
         &          & sunquake 2 &       &           &          &          & 7.1     & -343*   & /       \\
         &          & sunquake 3 &       &           &          &          & 21.0    & 236     & 235     \\
13       & 20140107 & M7.2       & 11944 & -13.6     & -13.0    & 10:12:00 & 22.1    & 436     & 680     \\
14       & 20140611 & X1.0       & 12087 & -66.0     & -18.0    & 09:03:00 & 29.5    & /       & 1338    \\
15       & 20141022 & M8.7       & 12192 & -23.4      & -13.0    & 01:39:45 & 8.8     &         &         \\
         &          & sunquake 1 &       &           &          &          & 6.1     & 133*    & -1   \\
         &          & sunquake 2 &       &           &          &          & 2.7     & 263     & 167     \\
16       & 20141022 & X1.6       & 12192 & -13.0     & -14.0    & 14:06:45 & 2.7     & 96*   & -41   \\
17     & 20141220 & X1.8       & 12242 & 29.3      & -18.5    & 00:20:15 & 5.6     &         &         \\
         &          & sunquake 1 &       &           &          &          & 2.7     & 7*    & 6*   \\
         &          & sunquake 2 &       &           &          &          & 2.9     & /     & 175*     \\
18       & 20150311 & X2.1       & 12297 & -22.0     & -17.0    & 16:17:15 & 12.9    & -315    & 36    \\
19       & 20150625 & M7.9       & 12371 & 40.3      & 12.6      & 08:15:00 & 29.6    &         &         \\
         &          & sunquake 1 &       &           &          &          & 14.7    & -372*    & /       \\
         &          & sunquake 2 &       &           &          &          & 14.9    & /       & 158     \\
20       & 20150928 & M7.6       & 12422 & 28.0      & -20.0    & 14:57:00 & 52.8    &         &         \\
         &          & sunquake 1 &       &           &          &          & 21.7    & 256     & 705     \\
         &          & sunquake 2 &       &           &          &          & 23.5    & 246     & /       \\
         &          & sunquake 3 &       &           &          &          & 7.5     & -112    & -372    \\
21       & 20170906 & X2.2       & 12673 & 33.1      & -9.1    & 09:04:30 & 13.8    &         &         \\
         &          & sunquake 1 &       &           &          &          & 7.0     & 276*    & 86*     \\
         &          & sunquake 2 &       &           &          &          & 6.8     & 56      & 21      \\
22       & 20170906 & X9.3$^\dagger$       & 12673 & 34.7      & -9.1    & 11:57:00 & 203.3   &         &         \\
         &          & sunquake 1 &       &           &          &          & 39.5    & 932     & 1075    \\
         &          & sunquake 2 &       &           &          &          & 34.5    & 645     & 189*    \\
23       & 20170907 & M7.3       & 12673 & 46.3      & -7.3    & 10:15:45 & 42.1    & 534     & 344     \\
24       & 20170907 & X1.3       & 12673 & 49.1      & -9.1    & 14:33:45 & 58.4    &         &         \\
         &          & sunquake 1 &       &           &          &          & 34.1    & -142*   & 19      \\
         &          & sunquake 2 &       &           &          &          & 13.0    & 400*    & 224     \\
         &          & sunquake 3 &       &           &          &          & 11.6    & /       & 99    \\
\hline
\end{tabular}

\raggedright $^\dagger$ Note that this flare generates more than two sunquake kernels. Due to the very complicated nature of this event, and in order not to allow one flaring event to give too much weight in this statistical study, we only list two strongest sunquake sources for this flare.
\label{stats}
\end{table}

We also examine the relationship between HMI continuum-intensity emissions and sunquakes, and between  Doppler transients and sunquakes. For the 60 flares surveyed, 49 have HMI continuum-intensity enhancements and 48 have Doppler transients, with 47 of them having both. Of the 24 flares that are sunquake active, all have both HMI continuum emissions and Doppler transients.  Flares with Doppler transients or HMI continuum emissions not necessarily generate sunquakes.  When there are sunquakes, their sources overlap with part of the Doppler transients. Similar relations were found by \citet{Bui15} (note their flare samples include more weak flares than~ours).

\section{Discussion}
\label{sec_discuss}

In this study, we develop a new method to derive observation-based Green's function for the helioseismic holography technique, and reconstruct the oscillatory velocity fields for 60 flare-hosting active regions. We find that 24 of the 60 studied flares gave a total of 41 sunquake events, with the emitted sunquake energies between a wide range of $10^{27} - 10^{29}$~erg. We examine the average oscillatory velocity at the sunquake sources during the flares' impulsive phases, which are determined from the photospheric observations in the sunquake sources, and also analyze the spatial relation between the sunquakes sources and flares. Of the 41 sunquake events, 25 of 31 (80.6\%) sunquakes in the $3-5$ mHz frequency band have average downward velocities during the flares' impulsive phases, and 33 of 38 (86.8\%) sunquakes in the $5-7$ mHz frequency band have net downward velocities. This is consistent with our hypothesis that sunquakes tend to occur when the downward flare impact strengthens a downward background oscillation. 
{For those events that do not show downward velocities at or near the flaring sites, we find that they happen rather randomly, without a correlation with the flares’ strength or flaring locations.} 
As for the spatial relation, a majority of sunquakes occur on or near the PILs; and the sunquake sources also tend to be located between or at the edge of the flare ribbons denoted by the HMI line-core intensity enhancements. All the sunquake-generating flares also have both continuum-intensity enhancements and Doppler transients in our survey, but flares with Doppler transients or continuum emissions do not necessarily generate sunquakes.

The use of observation-based Green's functions in this analysis improves the holography egression calculation. The Green's functions calculated explicitly for waves traveling out from sunspot umbra, penumbra, and near-sunspot quiet regions correct the travel time deficits introduced by sunspots, regardless of the causes of the deficits, and are also more realistic in the frequency dispersion and energy dissipation. Therefore, we believe our results are more reliable in the temporal determination of sunquakes than most previous studies. We notice, however, that the temporal resolution is limited by the frequency bandwidth, $2-7$ mHz, within which oscillatory signals are reconstructed in this study. A finer resolution requires an inclusion of frequencies below 2 mHz, but it is difficult to obtain with a good signal-to-noise ratio due to the low oscillatory power and strong convective power. However, that effect does not influence our results on the sign (direction) of the oscillatory velocity that is studied during the flare's impulsive phase.

\begin{table}[!htb]
\caption{Frequency of sunquake-associated flares in each flare class.}
\begin{tabular}{|c|c|c|c|}
\hline

Flare class & Sunquake-associate & Non-sunquake-associate & Frequency \\ \hline
$\leqslant$M7          & 4                  & 13                      & 0.24      \\ \hline
M8          & 1                  & 3                      & 0.25      \\ \hline
M9          & 1                  & 2                      & 0.33      \\ \hline
X1          & 11                 & 13                     & 0.46      \\ \hline
X2          & 4                  & 3                      & 0.57      \\ \hline
$\geqslant$X3          & 3                  & 2                      & 0.60      \\ \hline
\end{tabular}

\label{Freq}
\end{table}

Our hypothesis considers the photospheric background oscillation as a factor for a sunquake to occur, which was usually neglected in previous studies. Given a momentum injected into the photosphere during a flare, regardless of by shock waves, energetic particles, or Lorenz force, the background oscillation may get enhanced or diminished by the momentum injection in a stochastic sense depending on whether they are in the same direction, and this influences the flare's probability of generating detectable sunquakes. 
The combined velocity or momentum needs to stand out from the ambient background oscillations to become a detectable sunquake event.  
Given that the background oscillatory velocity at the time of flare impact is stochastically distributed, it is theoretically a stochastic process for the combined velocity to be larger than a certain threshold at which it is considered detectable relative to background. And its probability, $\bf P$(~flare impulse + background momentum $>$ threshold ), depends partly on the flare impulse. 
The flare impulse increases approximately with flare class, and so should be the probability for a flare to cause sunquakes. Table~\ref{Freq} summarizes the numbers of flares that are sunquake-associated or not as well as the frequency of sunquake-associated flares for each flare class. It is shown that the frequency increases approximately with flare class, supporting our hypothesis. However, it is acknowledged that the sample size is not large enough for a reliable estimate of probability. 
On the other hand, assuming most of the impulses from the flares occurring above are downward, the combined photospheric velocity that is able to successfully cause a sunquake should be downward, too. Our statistical result in Section~\ref{sec_statistics} on the averaged oscillatory velocity during the flares' impulsive phases supports this hypothesis, too. 
Besides the statistics, it is also noticed that for certain active regions, e.g., ARs 11890 and 12673, each hosts a few sunquake-generating flares, with similar sunquake locations for different flares. It seems that certain sunspots are prone to sunquakes, possibly because their momentum impacts on the photosphere during flares tend to be so overwhelming that the background oscillation only plays a minimal role. The capability of generating large momentum impacts may depend on the magnetic configuration, and a certain configuration of field lines can create a ``magnetic lens" to focus energy on one particular location of the photosphere, as suggested by \citet{Gre17}. 

Despite that our proposed scenario is supported by statistics, it is also acknowledged that the background oscillation considered works only as a selection rule that restricts the chance for a sunquake to occur, but does not explain how or when a flare triggers a sunquake in the first place. However, our comprehensive survey of strong flares in Solar Cycle 24, with reconstructed oscillatory velocity field, provides valuable materials for studying flares and sunquake-triggering mechanisms. The location and timing of the sunquakes can be  compared with UV/EUV, soft and hard X-ray, and $\gamma$-ray observations to further study the triggering mechanism  of sunquakes as well as the impacts of energetic particles on the photosphere. These data can also be combined with other photospheric observations, such as vector magnetograms, to study the magnetic imprints from the eruptive events in the corona \citep{Sun17}. 

Our analysis shows that at the sunquake source, the oscillatory velocity is primarily downward during the impulsive phase of the localized flaring; however, to fully support the hypothesis proposed in Section 1, it would be natural to examine if it is the case that the non-sunquake-generating flares do not occur above a downward-oscillating area. However,  another fact prevents us from performing such an analysis. As can be seen in Figures~\ref{all_M}~and~\ref{all_Ic}, a sunquake source is often a very compact region, located close to a flare ribbon that in contrast covers a rather large area. It is thus impractical to examine the averaged velocities for those non-sunquake-generating flares, because it is unclear which area and what size of the area we need to investigate. 

Moreover, we also recognize that due to the limited frequency bandwidth used in reconstructing the velocity field, the whole envelope of the oscillatory signals at the sunquake source gets substantially expanded. The result is that the sunquakes seem to start before the flares even start (Figure~\ref{curves_fig}). Although the oscillatory directions during the flares' impulsive phases do not get altered, the oscillatory velocity reconstructed for the period before the flare's impulsive phases is a combination of the background oscillations and the sunquake signals leaked into this time period. Therefore, this fact weakens our attempt of examining  accurately oscillations before the flare onset. However, it is difficult to assess how much the signals are altered. 

In summary, our statistics supports a sunquake selection rule that the coincidence of a strong flare's impulsive phase with a downward background oscillatory velocity near the flare's impact sites likely improves the chance of generating a detectable sunquake event. This may help us  to understand why some flares trigger sunquakes and others do not, and why strong flares are not necessarily accompanied by sunquakes.  More case studies can be done on the sunquake active or non-active flares listed in our survey to study the triggering mechanism of sunquakes. 


\acknowledgments {\it SDO} is a NASA mission, and HMI project is supported 
by the NASA contract NAS5-02139 to Stanford University. R.C. is partly 
supported by the NASA Earth and Space Science Fellowship NNX15AT08H.

\end{document}